\documentstyle[preprint,aps]{revtex}
\begin{document}
\draft
\title{Higher Dimensional Dilaton Black Holes with Cosmological Constant}
\author{Chang Jun Gao$^{1}$\thanks{E-mail: gaocj@mail.tsinghua.edu.cn}
Shuang Nan Zhang$^{1,2,3,4}$\thanks{E-mail:
zhangsn@mail.tsinghua.edu.cn}}
\address{$^{1}$Department of Physics and Center for Astrophysics, Tsinghua University, Beijing 100084, China(mailaddress)}
\address{$^2$Physics Department, University of Alabama in Huntsville, AL 35899, USA }
\address{$^3$Space Science Laboratory, NASA Marshall Space Flight Center, SD50, Huntsville, AL 35812, USA }
\address{$^4$Laboratory for Particle Astrophysics, Institute of High Energy Physics, Chinese Academy of Sciences, Beijing 100039, China}

\date{\today}
\maketitle

\begin{abstract}
\hspace*{7.5mm}The metric of a higher-dimensional dilaton black
hole in the presence of a cosmological constant is constructed. It
is found that the cosmological constant is coupled to the dilaton
in a non-trivial way. The dilaton potential with respect to the
cosmological constant consists of three Liouville-type potentials.
\end{abstract}

\section{introduction}
 \hspace*{7.5mm}There has been much interest in recent years in
 the dilaton gravity. It is of great importance to investigate how the properties of black holes are
modified when the dilaton field is present. Exact solutions of
charged dilaton black holes have been constructed by many authors.
It is found that dilaton changes the causal structure of the black
hole and leads to the curvature singularities at finite radii
[1-7]. These black holes are all asymptotically flat. In the
presence of one Liouville-type potential which is regarded as the
generalization of the cosmological constant, a class of charged
black hole solutions have been discovered [8, 9]. Unfortunately,
these solutions are asymptotically neither flat nor (anti)-de
Sitter.\\
\hspace*{7.5mm}In fact, Poletti, Wiltshire and Okai [10, 11, 12]
have shown that with the exception of a pure cosmological
constant, no asymptotically flat, asymptotically de Sitter or
asymptotically ant-de Sitter static spherically symmetric
solutions to the field equations associated with only one
Liouville-type potential exist. In a recent work, we obtained the
asymptotically de Sitter and asymptotically anti-de Sitter dilaton
solutions in four dimensions [13]. It is found that the dilaton
potential which is regarded as an extension of the cosmological
constant have three Liouville-type potentials. In this letter, we
extend it to arbitrary dimensions.
\section{Higher Dimensional Dilaton Black Holes with Cosmological Constant}
 \hspace*{7.5mm}We consider the n-dimensional theory in which
 gravity is coupled to dilaton and Maxwell field with an action
\begin{eqnarray}
S=\int{d^nx\sqrt{-g}\left[R-\frac{4}{n-2}\partial_{\mu}\phi\partial^{\mu}\phi-V\left(\phi\right)
-e^{-\frac{4\alpha\phi}{n-2}}F^2\right]},
\end{eqnarray}
where $R$ is the scalar curvature, $F^2=F_{\mu\nu}F^{\mu\nu}$ is
the usual Maxwell contribution, and $V\left(\phi\right)$ is a
potential of dilaton $\phi$ which is with respect to the
cosmological constant. $\alpha$ is an arbitrary constant governing
the strength of the
coupling between the dilaton and the Maxwell field.\\
\hspace*{7.5mm}Varying the action with respect to the metric,
Maxwell, and dilaton fields, respectively, yields
\begin{equation}
R_{\mu\nu}=\frac{4}{n-2}\left(\partial_{\mu}\phi\partial_{\nu}\phi+\frac{1}{4}
g_{\mu\nu}V\right)+2e^{-\frac{4\alpha\phi}{n-2}}\left(F_{\mu\alpha}
F_{\nu}^{\alpha}-\frac{1}{2n-4}g_{\mu\nu}F^2\right),
\end{equation}
\begin{equation}
\partial_{\mu}\left(\sqrt{-g}e^{-\frac{4\alpha\phi}{n-2}}F^{\mu\nu}\right)=0,
\end{equation}
\begin{equation}
\partial_{\mu}\partial^{\mu}\phi=\frac{n-2}{8}\frac{\partial V}{\partial \phi}-\frac{\alpha}
{2}e^{-\frac{4\alpha\phi}{n-2}}F^2.
\end{equation}
\hspace*{7.5mm}We choose the most general form of the metric for
the static dilaton black hole with a cosmological constant as
follows
\begin{equation}
ds^2=-U\left(r\right)dt^2+\frac{1}{U\left(r\right)}dr^2+f\left(r\right)^2d\Omega_{n-2}^2,
\end{equation}
where $r$ denotes the radial variable. Then the Maxwell equation
Eq.(3) can be integrated to give
\begin{equation}
F_{01}=\frac{Qe^{\frac{4\alpha\phi}{n-2}}}{f^{n-2}},
\end{equation}
where $Q$ is the electric charge of the black hole. With the
metric Eq.(5) and the Maxwell field Eq.(6), the equations of
motion Eqs.(2-4) reduce to three independent equations
\begin{equation}
\frac{1}{f^{n-2}}\frac{d}{dr}\left(f^{n-2}U\frac{d\phi}{dr}\right)=\frac{n-2}{8}\frac{\partial
V}{\partial \phi}+\alpha
\frac{Q^2e^{\frac{4\alpha\phi}{n-2}}}{f^{2n-4}},
\end{equation}
\begin{equation}
\frac{1}{f}\frac{d^2f}{dr^2}=-\frac{4}{\left(n-2\right)^2}\left(\frac{d\phi}{dr}\right)^2,
\end{equation}
\begin{equation}
\frac{1}{f^{n-2}}\frac{d}{dr}\left[U\frac{d}{dr}\left(f^{n-2}\right)\right]=
\frac{\left(n-2\right)\left(n-3\right)}{f^2}-V-2\frac{Q^2e^{\frac{4\alpha\phi}{n-2}}}{f^{2n-4}}.
\end{equation}
\hspace*{7.5mm}Since we have no knowledge of how the cosmological
constant is coupled to the dilaton, there are four unknown
quantities $U(r), f(r), \phi(r)$ and $V(\phi)$ in above three
equations of motion. We can not solve them in the usual way. So
let's turn our attention to the n-dimensional dilaton black
hole solution without the cosmological constant.\\
\hspace*{7.5mm}The metric for the well-known n-dimensional dilaton
black hole without the cosmological constant is given by [14]
\begin{eqnarray}
d{s}^2&=&-\left[1-\left(\frac{r_{+}}{{r}}\right)^{n-3}\right]
\left[1-\left(\frac{r_{-}}{{r}}\right)^{n-3}\right]^{1-\gamma\left(n-3\right)}
d{t}^2\nonumber\\&&+\left[1-\left(\frac{r_{+}}{{r}}\right)^{n-3}\right]^{-1}
\left[1-\left(\frac{r_{-}}{{r}}\right)^{n-3}\right]^{\gamma-1}d{r}^2\nonumber\\&&+
r^2\left[1-\left(\frac{r_{-}}{r}\right)^{n-3}\right]^{\gamma}d\Omega_{n-2}^2,
\end{eqnarray}
where $r_{+}$, $r_{-}$ are two event horizons of the black hole
and $\gamma$ is a constant which is related to the coupling
constant $\alpha$ and the dimensions $n$ of the spacetime. Here $r$ denotes the radial variable. \\
\hspace*{7.5mm}Compare Eq.(5) with Eq.(10), we find that
$g_{00}\neq -g^{11}$ in Eq.(10) which is different from Eq.(5). To
achieve $g_{00}=-g^{11}$, we can rewrite the metric Eq.(10) from
the $(t, r)$ coordinate system to the $(t, x)$ coordinate system
via the following coordinates transformation
\begin{eqnarray}
x=\int dr
\left[1-\left(\frac{r_{-}}{{r}}\right)^{n-3}\right]^{-\gamma\left(n-4\right)/2},\
\ \ \ i.e.,\ \ \ \
r^{'}=\left[1-\left(\frac{r_{-}}{{r}}\right)^{n-3}\right]^{\gamma\left(n-4\right)/2},
\end{eqnarray}
namely, the radial variable $r$ is replaced by $x$. Here the prime
${'}$ denotes the derivative with respect to $x$. Then Eq.(10) is
reduced to the following form
\begin{eqnarray}
d{s}^2&=&-\left\{1-\left[\frac{r_{+}}{{r\left(x\right)}}\right]^{n-3}\right\}
\left\{1-\left[\frac{r_{-}}{{r\left(x\right)}}\right]^{n-3}\right\}^{1-\gamma\left(n-3\right)}
d{t}^2\nonumber\\&&+\left\{1-\left[\frac{r_{+}}{{r\left(x\right)}}\right]^{n-3}\right\}^{-1}
\left\{1-\left[\frac{r_{-}}{{r\left(x\right)}}\right]^{n-3}\right\}^{-1+\gamma\left(n-3\right)}d{x}^2\nonumber\\&&+
r\left(x\right)^2\left\{1-\left[\frac{r_{-}}{{r\left(x\right)}}\right]^{n-3}\right\}^{\gamma}d\Omega_{n-2}^2,
\end{eqnarray}
where the function $r\left(x\right)$ is determined by Eq.(11). It ia apparent $g_{00}=-g^{11}$ in Eq.(12).\\
\hspace*{7.5mm}Inspecting the four-dimensional dilaton black hole
solution with a cosmological constant [13]
\begin{eqnarray}
d{s}^2&=&-\left[\left(1-\frac{r_{+}}{{x}}\right)\left(1-\frac{r_{-}}{{x}}\right)^{\frac{1-\alpha^2}{1+\alpha^2}}
-\frac{1}{3}\lambda
x^2\left(1-\frac{r_{-}}{x}\right)^{\frac{2\alpha^2}{1+\alpha^2}}\right]
d{t}^2\nonumber\\&&+\left[\left(1-\frac{r_{+}}{{x}}\right)\left(1-\frac{r_{-}}{{x}}\right)^{\frac{1-\alpha^2}{1+\alpha^2}}
-\frac{1}{3}\lambda
x^2\left(1-\frac{r_{-}}{x}\right)^{\frac{2\alpha^2}{1+\alpha^2}}\right]^{-1}d{x}^2\nonumber\\&&+
x^2\left(1-\frac{r_{-}}{x}\right)^{\frac{2\alpha^2}{1+\alpha^2}}d\Omega_{2}^2,
\end{eqnarray}
where $\lambda$ is the cosmological constant, we suppose the
n-dimensional dilaton black hole solution with a cosmological
constant has the following form
\begin{eqnarray}
&&U=\left\{1-\left[\frac{r_{+}}{{r\left(x\right)}}\right]^{n-3}\right\}
\left\{1-\left[\frac{r_{-}}{{r\left(x\right)}}\right]^{n-3}\right\}^{1-\gamma\left(n-3\right)}
-\frac{1}{3}\lambda r\left(x\right)^2\left\{1-\left[\frac{r_{-}}{{r\left(x\right)}}\right]^{n-3}
\right\}^{\gamma},\nonumber\\
&&f=r\left(x\right)\left\{1-\left[\frac{r_{-}}{{r\left(x\right)}}\right]^{n-3}\right\}^{\gamma/2}.
\end{eqnarray}
In the new coordinate system $(t, x)$, the equations of motion
Eqs.(7-9) become
\begin{equation}
\frac{1}{f^{n-2}}r^{'}\frac{d}{dr}\left(f^{n-2}Ur^{'}\frac{d\phi}{dr}\right)=\frac{n-2}{8}\frac{\partial
V}{\partial \phi}+\alpha
\frac{Q^2e^{\frac{4\alpha\phi}{n-2}}}{f^{2n-4}},
\end{equation}
\begin{equation}
\frac{1}{f}\frac{d}{dr}\left(r^{'}\frac{df}{dr}\right)=-\frac{4}{\left(n-2\right)^2}\left(\frac{d\phi}{dr}\right)^2r^{'},
\end{equation}
\begin{equation}
\frac{1}{f^{n-2}}r^{'}\frac{d}{dr}\left[Ur^{'}\frac{d}{dr}\left(f^{n-2}\right)\right]=
\frac{\left(n-2\right)\left(n-3\right)}{f^2}-V-2\frac{Q^2e^{\frac{4\alpha\phi}{n-2}}}{f^{2n-4}},
\end{equation}
where $r$ denotes $r(x)$.\\
\hspace*{7.5mm}Substituted the expressions of $r^{'}$ and $f$ into
Eq.(16), the dilaton field $\phi$ is obtained
\begin{eqnarray}
e^{2\phi}&=&e^{2\phi_0}\left[1-\left(\frac{r_{-}}{r}\right)^{n-3}\right]
^{\left(n-2\right)\sqrt{\gamma}\sqrt{2+3\gamma-n\gamma}/2},
\end{eqnarray}
where $\phi_0$ is the integration constant which has the meaning
of the asymptotic value of dilaton.\\
\hspace*{7.5mm}We in the first place obtain the potential of
dilaton $V(\phi)$ from Eq.(17) and then insert both $V(r)$ and
$\phi(r)$ into Eq.(15), and then after a lengthy calculation, we
find that Eq.(15) is satisfied when
\begin{eqnarray}
\gamma &=&\frac{2\alpha^2}{\left(n-3\right)\left(n-3+\alpha^2\right)},\nonumber\\
Q^2&=&\frac{\left(n-2\right)\left(n-3\right)^2}{2\left(n-3+\alpha^2\right)}
e^{-\frac{4\alpha\phi_0}{n-2}}r_+^{n-3}r_-^{n-3},
\end{eqnarray}
and then the dilaton potential can be written as
\begin{eqnarray}
V\left(\phi\right)&=&\frac{\lambda}{3\left(n-3+\alpha^2\right)^2}
\left[-\alpha^2\left(n-2\right)\left(n^2-n\alpha^2-6n+\alpha^2+9\right)
e^{-\frac{4\left(n-3\right)\left(\phi-\phi_0\right)}{\left(n-2\right)\alpha}}\right.\nonumber
\\ &&\left. +\left(n-2\right)
\left(n-3\right)^2\left(n-1-\alpha^2\right)e^{\frac{4\alpha\left(\phi-\phi_0\right)}{n-2}}\right.\nonumber
\\ &&\left. +4\alpha^2\left(n-3\right)
\left(n-2\right)^2e^{\frac{-2\left(\phi-\phi_0\right)\left(n-3-\alpha^2\right)}{\left(n-2\right)\alpha}}\right].
\end{eqnarray}
\hspace*{7.5mm}It is clear the cosmological constant is coupled to
the dilaton in a very non-trivial way. When $n=4$, Eqs.(19-20)
recover our previous results [13].\\
\hspace*{7.5mm}Come back to the $(t, r)$ coordinate system, we
present the metric of a dilaton black hole with the cosmological
constant
\begin{eqnarray}
d{s}^2&=&-\left\{\left[1-\left(\frac{r_{+}}{{r}}\right)^{n-3}\right]
\left[1-\left(\frac{r_{-}}{{r}}\right)^{n-3}\right]^{1-\gamma\left(n-3\right)}
-\frac{1}{3}\lambda
r^2\left[1-\left(\frac{r_{-}}{{r}}\right)^{n-3} \right]^{\gamma}
\right\}d{t}^2\nonumber\\&&+\left\{\left[1-\left(\frac{r_{+}}{{r}}\right)^{n-3}\right]
\left[1-\left(\frac{r_{-}}{{r}}\right)^{n-3}\right]^{1-\gamma\left(n-3\right)}
-\frac{1}{3}\lambda
r^2\left[1-\left(\frac{r_{-}}{{r}}\right)^{n-3} \right]^{\gamma}
\right\}^{-1}\nonumber
\\ &&\cdot\left[1-\left(\frac{r_{-}}{{r}}\right)^{n-3}
\right]^{-\gamma\left(n-4\right)}d{r}^2+
r^2\left[1-\left(\frac{r_{-}}{{r}}\right)^{n-3}\right]^{\gamma}d\Omega_{n-2}^2.
\end{eqnarray}
It is apparent that this spacetime is asymptotically
(anti)-de-Sitter. When $\lambda=0$, it restores to the result of
Horowitz and Strominger [14]. When $n=4$, it restores to our
previous result [13]. When the coupling constant $\alpha=0$, i.e.,
$\gamma=0$, it restores to the well-known n-dimensional
Reissner-Nordstrom-de
Sitter metric.\\
\hspace*{7.5mm}In conclusion, we have constructed the
higher-dimensional dilaton black hole solution in the presence of
the cosmological constant. So far no fully satisfactory de Sitter
version of black hole solutions in string theory has been found
[15]. Our solution is asymptotically both flat and (anti)-de
Sitter. Thus it may be a fully satisfactory solution of string
theory. We found that the dilaton potential with respect to the
cosmological constant is not of the form of a pure cosmological
constant but the form of three-Liouville-type potential. This is
different from the Einstein-Maxwell theory.
 \hspace*{7.5mm}\acknowledgements This study is
supported in part by the Special Funds for Major State Basic
Research Projects and by the National Natural Science Foundation
of China. SNZ also acknowledges supports by NASA's Marshall Space
Flight Center and through NASA's Long Term Space Astrophysics
Program.

\end{document}